\documentclass[twocolumn,aps,prl] {revtex4} 
\usepackage{graphicx}
\begin{document}
\pagestyle{empty} 
\title{On the elastic energy and stress correlation in the contact between elastic solids with
randomly rough surfaces}
\author{  
B.N.J. Persson} 
\affiliation{IFF, FZ-J\"ulich, 52425 J\"ulich, Germany}

\begin{abstract}
When two elastic solids with randomly rough surfaces are brought in contact, a 
very inhomogeneous stress distribution $\sigma({\bf x})$ will occur at the interface.
Here I study the elastic energy and the 
correlation function $\langle \sigma({\bf q})\sigma({\bf -q})\rangle$,
where $\sigma({\bf q})$ is the Fourier transform of $\sigma({\bf x})$ and 
where $\langle ..\rangle$ stands for ensemble average. 
I relate $\langle \sigma({\bf q})\sigma({\bf -q})\rangle$
to the elastic energy stored at the interface, and I show that for self affine fractal
surfaces, quite generally $\langle \sigma({\bf q})\sigma({\bf -q})\rangle \sim q^{-(1+H)}$,
where $H$ is the Hurst exponent of the self-affine fractal surface.
\end{abstract}
\maketitle


The contact between elastic solids with surfaces with roughness on many length scales
is a topic of great practical importance. Particularly important are the 
area of real contact $A$, which in most cases determines the sliding friction force, 
and the asperity induced elastic energy $U_{\rm el}$ stored at the interface, which is directly
related to the interfacial separation\cite{PerssonPRL}.

In a recent paper Campana, M\"user and Robbins\cite{Carlos} have presented numerical simulation results 
for the stress distribution at the interface between two elastic solids with randomly
rough (self affine fractal)
surfaces. They have shown that the stress correlation function
$$\langle \sigma({\bf q}) \sigma (-{\bf q}) \rangle \sim q^{-(1+H)}$$
where $H$ is the Hurst exponent of the self affine fractal surface. Here I show that this
is also the prediction of the contact mechanics theory of Persson\cite{JCPpers}. I also derive
some other results related to the elastic energy and the accuracy of the Persson theory. 

Consider the contact between two elastic solids with rough surfaces, but which appear flat at
low resolution. 
We can write the elastic energy stored in the vicinity of the asperity contact regions as
$$U_{\rm el} = {1\over 2} \int d^2x \ \langle \sigma({\bf x}) u({\bf x}) \rangle\eqno(1)$$
where $u({\bf x})$ and $\sigma({\bf x})$ are the normal displacement and the normal stress,
respectively.
We write 
$$\sigma({\bf x}) = \int d^2q \ \sigma({\bf q}) e^{i{\bf q}\cdot {\bf x}}\eqno(2)$$
and similar for $u({\bf x})$. Substituting this in (1) gives
$$U_{\rm el} = (2\pi )^2 {1\over 2} \int d^2q \ \langle \sigma({\bf q}) u(-{\bf q}) \rangle\eqno(3)$$
Next, using that\cite{JCPpers}
$$\sigma({\bf q}) = {1\over 2} E^* q u({\bf q})\eqno(4)$$
gives
$$U_{\rm el} = (2\pi )^2 {1\over E^* } \int d^2q \ q^{-1} \langle \sigma({\bf q}) 
\sigma(-{\bf q}) \rangle\eqno(5)$$
and
$$U_{\rm el} = (2\pi )^2 {E^* \over 4} \int d^2q \ q \langle u({\bf q}) 
u(-{\bf q}) \rangle\eqno(6)$$
For complete contact $u({\bf q})=h({\bf q})$ and using the definition
$$\langle h({\bf q}) h(-{\bf q}) \rangle= {A_0 \over (2\pi )^2} C(q)\eqno(7)$$
gives for complete contact
$$U_{\rm el} = {E^* A_0 \over 4} \int d^2q \ q C(q)\eqno(8)$$
Now, for incomplete contact I define $W(q)$ so that
$$U_{\rm el} = {E^* A_0 \over 4} \int d^2q \ q C(q) W(q)\eqno(9)$$
I have argued elsewhere\cite{P1} that $W(q)=P(q)$ is the relative contact area when the interface is studied
at the magnification $\zeta=q/q_0$. 
The qualitative explanation is that the solids will 
deform mainly in the regions
where they make contact and most of the elastic energy will arise from the contact regions.
In particular, for complete contact $P(q)=1$ and in this limit using $W=P=1$ would be exact. 
I will show below that $W(q)=P(q)$ follows directly from the contact mechanics theory
of Persson, but for the moment
we simply {\it define} $W(q)$ via (9). 

Comparing (5) and (9) gives
$$\langle \sigma({\bf q}) \sigma (-{\bf q}) \rangle =  
\left ( {E^* \over 4 \pi}\right )^2 A_0 q^2 C(q) W(q)\eqno(10)$$
In what follows we need
$$\int d^2q \ \langle \sigma ({\bf q})\sigma (-{\bf q})\rangle$$
$$ = {1\over (2\pi )^4}\int d^2x d^2x' 
\int d^2q \ \langle \sigma({\bf x})\sigma ({\bf x'})\rangle e^{-i{\bf q}\cdot ({\bf x}-{\bf x'})}$$
$$={1\over (2\pi)^2}\int d^2x \ \langle \sigma^2 ({\bf x}) \rangle = {A_0\over (2\pi)^2}\langle 
\sigma^2 \rangle\eqno(11)$$
Substituting (10) in (11) gives
$$\int_{q_0}^q dq \ q^3 C(q) W(q)= {2\over \pi} {\langle 
\sigma^2 \rangle\over {E^*}^2}\eqno(12)$$
All the equations (1)-(12) presented above are exact. 

Now, it is easy to calculate $\langle \sigma^2 \rangle $ approximately 
using the Persson contact
mechanics theory\cite{JCPpers}. The basic equation for the stress distribution $P(\sigma, \zeta)$
$${\partial P \over \partial \zeta} =  f(\zeta) {\partial^2 P \over \partial \sigma^2}\eqno(13)$$
where
$$f(\zeta)= {\pi \over 4} {E^*}^2 q_0 q^3 C(q)\eqno(14)$$
where $q=q_0\zeta$. 
Multiply (13) with $\sigma^2$ and integrate over $\sigma$:
$${d \langle \sigma^2 \rangle \over d \zeta} = 
f(\zeta) \int_0^\infty \sigma^2 {\partial^2 P \over \partial \sigma^2}$$
$$=2 f(\zeta) \int_0^\infty d\sigma P(\sigma,\zeta)\eqno(15)$$
where I have performed two partial integrations. Now, note that
$$\int_0^\infty d\sigma P(\sigma,\zeta) = {A(\zeta)\over A_0}= P(\zeta)\eqno(16)$$
which I also denote as $P(q)$ ($q=q_0\zeta$) for simplicity.
Using the definition of $f(\zeta)$ we get
$$\langle \sigma^2 \rangle = {\pi \over 2}{E^*}^2 \int_{q_0}^q dq \ q^3C(q)P(q)\eqno(17)$$ 
Substituting (17) in (12) gives
$$\int_{q_0}^q dq \ q^3 C(q) W(q)
= \int_{q_0}^q dq \ q^3C(q)P(q)\eqno(18)$$
Thus we get $W(q)=P(q)$. The resulting equation for the
elastic energy was used in Ref. \cite{P1}, but only qualitative arguments was given for its
validity (see above). Here we have proved that it follows rigorously within
the Persson contact mechanics theory. Substituting this result in (10) gives
$$\langle \sigma({\bf q}) \sigma (-{\bf q}) \rangle =  
\left ( {E^* \over 4 \pi}\right )^2 A_0 q^2 C(q) P(q)\eqno(19)$$
Assume now a self affine fractal surface. In this case
$$C(q)\sim q^{-2(H+1)}\eqno(20)$$ 
If we assume that the relative contact area is small then
$$P(q)\approx [\pi G(q)]^{-1/2}\eqno(21)$$
where
$$G(q)=\left ({E^*\over \sigma_0}\right )^2 {\pi \over 4} \int_{q_0}^q dq \ q^3 C(q)$$
$$\sim q^{2(1-H)}-q_0^{2(1-H)} \approx q^{2(1-H)}\eqno(21)$$
when $q>>q_0$. Thus,
$$P(q) \sim q^{H-1}\eqno(22)$$
when the contact is small and $q>>q_0$. Substituting these results into (19) gives the $q$-dependence
$$\langle \sigma({\bf q}) \sigma (-{\bf q}) \rangle \sim q^{-(1+H)}\eqno(23)$$
in good agreement with recent numerical studies of contact between elastic solids with randomly
rough surfaces\cite{Carlos}.

Detailed analysis of Molecular Dynamics\cite{YangPersson}, Finite Element Method\cite{FEM}
and Green's function Molecular Dynamics\cite{Mus} calculations indicate some small deviations
from the prediction of the theory of Persson. Thus, the contact area calculated as a function of
the squeezing pressure tend to be about $\sim 20 \%$ larger at small pressures 
(where $A$ varies linearly with $p$)
than predicted by the Persson theory. When the squeezing pressure increases, the deviation becomes smaller
and vanish at complete contact. Similarly, the elastic energy in the asperity contact regions may be 
overestimated in the theory of Persson  
at low squeezing pressure (see \cite{YangPersson}), while the difference
decreases at higher squeezing pressures and vanishes for complete contact. Here we will use the theory
above to show that these two facts are in fact related. 

We focus on small contact where the area $A$ of contact depends linearly on the nominal squeezing
pressure $p$. Since Persson's theory predict smaller contact area than numerical simulations,
the pressure in the contact regions will be higher and we expect that the prediction of the Persson
theory for $\langle \sigma^2\rangle$ will be higher than observed in the numerical simulations.
In fact one can show that
$$\langle \sigma^2 \rangle \approx \langle \sigma^2 \rangle_{\rm P} A_{\rm P}/A\eqno(24)$$
where the index ``P'' refer to the Persson theory. Thus, using (12), which holds exactly,
we conclude that $W(q) \approx P(q) A_{\rm P}/A$. Thus, at low squeezing pressures, using
the Persson prediction $W(q)=P(q)$ will overestimate the elastic energy by a factor of
$A_{\rm P}/A$, i.e., by about $\sim 20 \%$.

\end{document}